\documentclass[aps,prb,reprint,twocolumn,amsmath,amssymb,showpacs,superscriptaddress]{revtex4-1}

\usepackage[english]{babel}
\usepackage{bm}
\usepackage{xcolor}
\usepackage{amsmath}
\usepackage{braket}
\usepackage{graphicx, float}
\usepackage{ulem} 
\definecolor{dkblue}{rgb}{0,0,0.4} 

\newcommand{\Hnet}{H_{\text{network}}}
\newcommand\dif{\mathop{}\!\mathrm{d}}
\newcommand{\llangle}{\langle\hskip -0.5ex \langle}
\newcommand{\rangll}{\rangle\hskip -0.5ex \rangle}
\newcommand{\expctt}[1]{\llangle #1 \rangll}
\usepackage[bookmarks=true,colorlinks,citecolor=blue,urlcolor=blue,linkcolor=blue]{hyperref}

\begin{document}

\title{Localization crossover and subdiffusive transport in a classical facilitated network model of a disordered, interacting quantum spin chain}

\author{K. Klocke}
\affiliation{Department of Physics, University of California, Berkeley, California 94720, USA}

\author{C. D. White}
\affiliation{Joint Center for Quantum Information and Computer Science, NIST/University of Maryland, College Park, Maryland 20742, USA}
\affiliation{Condensed Matter Theory Center, University of Maryland, College Park, Md, 20742}

\author{M. Buchhold}
\affiliation{Institut f\"ur Theoretische Physik, Universit\"at zu K\"oln, D-50937 Cologne, Germany}

\begin{abstract}
We consider the random-field Heisenberg model, a paradigmatic model for many-body localization (MBL), and add a Markovian dephasing bath
coupled to the Anderson orbitals of the model's non-interacting limit.
We map this system to a classical facilitated hopping model that is computationally tractable for large system sizes, and investigate its dynamics.
The classical model exhibits a robust crossover between an ergodic (thermal) phase and a frozen (localized) phase.
The frozen phase is destabilized by thermal subregions (bubbles), which thermalize surrounding sites by providing a fluctuating interaction energy and so enable off-resonance particle transport.
Investigating steady state transport, we observe that the interplay between thermal and frozen bubbles leads to a clear transition between diffusive and subdiffusive regimes.
This phenomenology both describes the MBL system coupled to a bath, and
provides a classical analogue for the many-body localization transition in the corresponding quantum model, in that the classical model displays long local memory times.
It also highlights the importance of the details of the bath coupling in studies of MBL systems coupled to thermal environments.
\end{abstract}

\maketitle

\section{Introduction}

Generic isolated interacting quantum systems are ergodic and evolve to local thermal equilibrium, regardless of the system's initial state.
This phenomenon has been broadly observed in numerical simulations and experiments, and is explained by the Eigenstate Thermalization Hypothesis (ETH)\cite{Deutsch_1991_ETH,Srednicki_1994_ETH,Tasaki_1998_ETH,Rigol_2008_ETH,dalessio_quantum_2016}.
Some interacting quantum systems, however, display non-ergodic dynamics.
Of particular interest are many-body localized (MBL) systems,
which are interacting many-body systems that become localized upon increasing disorder above a critical strength \cite{Basko_2006_MBL,Oganesyan_2007,Pal_2009,Altman_2015,Nandkishore_2015}.
MBL dynamics have been experimentally observed in cold-atom experiments in one dimension~\cite{Schreiber_2015,Bordia_2016},
and there is an ongoing debate over the existence of a stable MBL phase in higher dimensions.
The transition between an ergodic (thermalizing) phase and a non-ergodic (localized) phase falls outside the standard Landau paradigm of phase transitions and has therefore attracted much attention\cite{Luitz_2017_review, Agarwal_2017, Laflorencie_2020, Potter_2015, Vosk_2015, Khemani_2017, Zhang_2016,Morningstar_2019}.

\begin{figure}[t!]
    \centering
    \includegraphics[width=\columnwidth]{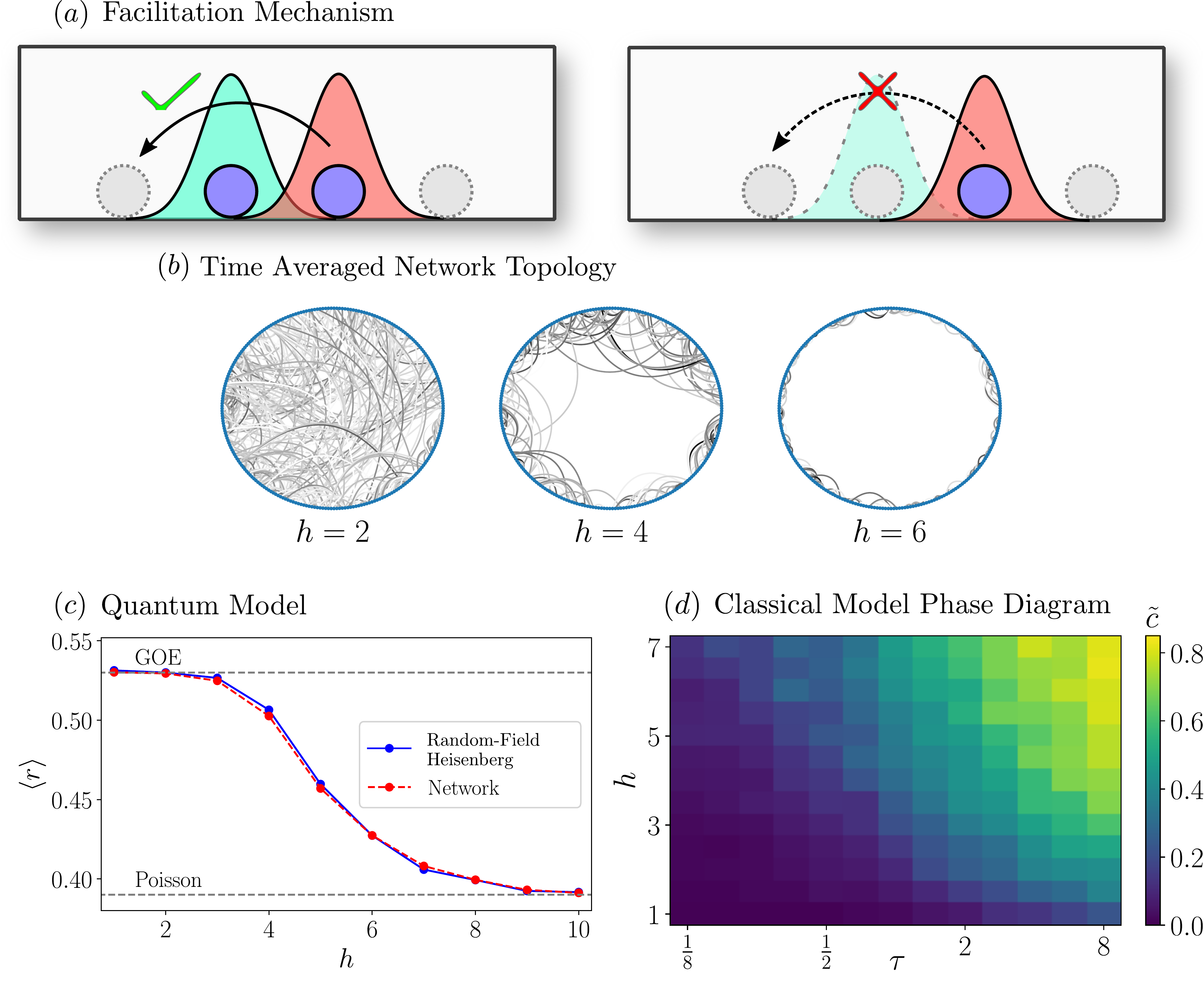}
    \caption{
    (a) Particle hopping in the facilitated network model requires the presence of a nearby occupied orbital (green) to facilitate the transition.
    (b) Transition in the network structure (via log-scaled time averaged link weights) from a highly connected graph at weak disorder to a sparse graph with several strongly connected components corresponding to thermal bubbles.
    (c) The level-statistics in the quantum network model reproduce the MBL transition in the random field Heisenberg model with high accuracy.
    (d) Phase diagram for the classical model, showing a transition in the fraction of frozen sites as a function of the disorder strength $h$ and dephasing time $\tau$.
    }
    \label{fig:summary_figure}
\end{figure}

A number of phenomenological pictures amenable to renormalization group treatment have been put forward to describe the transition~\cite{Potter_2015,Vosk_2015,dumitrescu_scaling_2017,goremykina_analytically_2018,dumitrescu_kosterlitz-thouless_2018,Morningstar_2019,morningstar_many-body_2020}.
These theories rest implicitly or explicitly on the notion that interactions destabilize an Anderson insulator by hybridizing the Anderson orbitals~\cite{Thiery_2018,thiery_microscopically_2017,de_roeck_stability_2017}.
Even deep in the localized phase,
rare regions of locally weak disorder bring nearby Anderson orbitals into resonance.
As the system approaches the MBL transition
these rare regions grow more common
and their influence expands via an ``avalanche'' process;
at the transition, this process results in resonant couplings throughout the system.
Conversely even when the system is ergodic, rare regions of locally strong disorder act as bottlenecks for transport and cause subdiffusion.
But the precise nature of the transition and the associated real-time dynamics remain challenging to study in a fully microscopic setting,
because numerical  methods are stymied by the development of delicate networks of resonating Anderson orbitals near the transition
\footnote{
Exact diagonalization studies are limited by strong finite size effects~\cite{chandran_finite_2015}.
Matrix product state methods are effective deep in the MBL phase.
There the slow growth of entanglement~\cite{bardarson_unbounded_2012}
means that time evolution with
matrix product states~\cite{vidal_efficient_2003,vidal_efficient_2004,white_real-time_2004} can treat dynamics to long times,
and methods exist for targeting eigenstates~\cite{werner_many-body_2015,lim_many-body_2016,yu_finding_2017}.
But near the transition rapid entanglement growth challenges matrix product state methods
\cite{prosen_many-body_2008,chanda_time_2020}.
Deep in the ergodic phase methods built on matrix product operators,
which can efficiently represent Gibbs states~\cite{hastings_solving_2006,schuch_approximating_2015}, show promise:
a nonequilibrium steady state study\cite{znidaric_diffusive_2016}
found a clear transition from diffusive to subdiffusive behavior,
and a new generation of matrix product operator and related methods \cite{white_quantum_2017,rakovszky_dissipation-assisted_2020,kvorning_time-evolution_2021,white_effective_2021}
offer the prospect of well-motivated, accurate approximations for real-time evolution.
But these methods are ill-equipped to handle the slow equilibration times and nontrivial correlations that develop near the MBL transition.
}.

The difficulty of simulating networks of resonating Anderson orbitals,
combined with the avalanche picture of thermal inclusions acting as baths and prompting a delocalizing cascade
and the extensive literature on MBL systems coupled to baths~\cite{Fischer_2016,Levi_2016,Ponte_2017,Everest_2017,Luschen2017,Wu_2019,Wu_2019a,van_Nieuwenburg_2017,Crowley_2020},
suggest that one break these networks by adding baths in a way that is easy to simulate.
But is there a bath structure that is both easy to simulate and preserves key features of the MBL transition?
A bulk dephasing bath is accessible to matrix product operator simulations, but it results in diffusion after length- and time-scales determined by the bath coupling---even in regimes where the the model without bath coupling displays subdiffusion\cite{znidaric_dephasing_2017}.
A bath of this kind is not sensitive to the distinction between Anderson localization and many-body localization.
Even in the non-interacting limit a homogeneous bath will destroy the system's Anderson orbitals, resulting in diffusive transport.

In this work, we investigate what parts of the phenomenology of the MBL transition remain when we weakly couple an interacting, disordered spin chain to a bath via its non-interacting Anderson orbitals. The crucial property of such a bath is that it preserves the integrity of the Anderson orbitals;
in the non-interacting limit the system will still behave as an an Anderson insulator even in the presence of a bath.
This setup therefore specifically probes the interplay of interaction and homogeneous dephasing in destabilizing the Anderson insulator. Moreover, by tuning the bath coupling we can crudely probe the role of specifically resonant interactions.

We start from a disordered spin chain weakly coupled to a Markovian dephasing bath, and map it to a kinetically constrained classical model.
In the original quantum model, interactions have two effects: they give facilitated transitions between Anderson orbitals, and they renormalize the energies of those Anderson orbitals.
The coupling to the bath then induces a dephasing of the exponentially slow transition processes, which results in classical Markov dynamics: in a fermionized model, the fermions hop between sites at rates determined by the magnitude of the facilitated hopping term and the difference between configuration energies.
Crucially, this dephasing results in hopping between slightly off-resonant orbitals; the allowable energy difference can be tuned by the bath coupling. We simulate this classical model using a kinetic Monte Carlo scheme.

We find that the resulting system displays qualitatively different physics compared to systems with baths that destroy Anderson localization by coupling to single physical sites. In particular, we see many signatures of localization.
Systems where Anderson localization is broken by the bath are well-characterized by a single timescale polynomial in the bath coupling rate and the disorder width.\cite{Fischer_2016,Levi_2016}
Our system, in contrast, displays a crossover characterized by a diverging correlation time, indicating that---in the ``frozen'' regime---it has the ``memory" characteristic of MBL systems.
This diverging correlation time results from the formation of rare regions of anomalously slow dynamics, whose average size increases with the disorder strength.
These regions act as a bottleneck to particle transport: once their density has reached a significant value,  we see nonequilibrium steady state current scaling with length as \[ j \propto L^{-\alpha}\;, \] where $\alpha > 1$.
Unlike the local (single orbital) observables---for which there is a smooth crossover between the active and frozen phases---the particle current shows a clearer transition from diffusive to subdiffusive transport.
The onset of subdiffusion at sufficiently strong disorder reproduces an essential feature of the MBL transition.
We then propose a feedback mechanism for the bath bandwidth that, when the system is considered its own bath, may reproduce the appropriate critical disorder strength observed in the isolated quantum model.

The paper is organized as follows.
In Sec.~\ref{sec:quantum_model} we review the random-field Heisenberg model, and then
approximate it by a quantum facilitated hopping model on Anderson orbitals. We then describe the details and the effect of a bath coupled to those Anderson orbitals,
and map the resulting dynamics to a classical facilitated network model.
We present the results of kinetic Monte Carlo simulations of the classical model in Sec.~\ref{sec:results};
there we explore the freezing transition,
the presence of rare thermal bubbles,
and the onset of subdiffusion in the nonequilibrium steady state current.
Finally, in Sec.~\ref{ss:percolation} we strengthen the connection between the classical model and the quantum facilitated hopping model by identifying the freezing transition in the dynamics as a real-space projection of a percolation-type transition in the hypergraph defined over configuration space.

\section{Models: From the Random-field Heisenberg Chain to a Classical Facilitated Anderson Network}\label{sec:quantum_model}

Before we present the details of this section, we provide a brief summary: We start with a paradigmatic setup for MBL, the random-field Heisenberg model (Sec.~\ref{ss:rfheis}).
Then we rephrase this in the language of Anderson orbitals,
and keep the leading order and next to leading order interaction terms;
this results in a quantum facilitated hopping network (Sec.~\ref{ss:fac-net}).
In order to controllably probe the effect of resonant sub-networks of Anderson orbitals on transport,
we weakly couple the system's Anderson orbitals to a particle number conserving, i.e., dephasing, bath.
We take the bath to be Markovian and have a tunable coupling strength $\tau^{-1}$ (Sec.~\ref{ss:bath});
$\tau$ is then the dephasing time for the Anderson orbitals.

This dephasing time is the key parameter in our treatment of transport in resonant networks.
The bath gives incoherent transport between sites $i, j$ when the
difference $\Delta_{ij}$ in many-body energies is $|\Delta_{ij}| \lesssim \tau^{-1}$,
so by tuning $\tau$ we tune the energy window for (bath-induced) resonance between different Anderson sites.
The incoherent transport then leads naturally to a classical hopping model   (Sec.~\ref{ss:classical_model}).

\subsection{Random-Field Heisenberg Chain}\label{ss:rfheis}
The random-field Heisenberg model
is a paradigmatic model for many-body localization
\cite{Pal_2009,Luitz_2015,prosen_many-body_2008,Laflorencie_2020}.
When written with spin operators it describes a chain of spin-$\frac{1}{2}$ degrees of freedom with nearest neighbor interactions and random onsite fields.
Applying the Jordan-Wigner transformation gives fermions with nearest-neighbor hopping and interactions, as well as a random onsite energy:
if $f_l$ are fermion operators $\{f_m, f^\dagger_l\}=\delta_{l,m}$ and $n_l=f^\dagger_lf_l$
on a lattice site $l$,
the Hamiltonian is
\begin{equation}\label{eq:Heisenberg}
    H =  V\sum_l\left(f^\dagger_l f_{l+1} + f^\dagger_{l+1} f_{l} +  2Jn_l n_{l+1} +  h_l n_l\right)\;.
\end{equation}
The onsite energy $h_l$ is drawn uniformly from $h_l\in[-h,h]$.
We choose energy units so that $V=1$.
Throughout this work we take the dimensionless interaction parameter
$J = 0$ for a non-interacting model
or $J = 1$ for an interacting model.

The non-interacting model at $J = 0$ is called the Anderson model~\cite{Mirlin_review}.
It displays Anderson localization:
the energy eigenstates are (Slater determinants of)
single-particle Anderson orbitals
\begin{equation}
    c_j=\sum_l \psi^{(j)}_l f_l
\end{equation}
where each wavefunction $\psi^{(j)}$
is exponentially localized around site $j$ \cite{Dupont_2019}
\begin{equation}
    \psi^{(j)}_l\sim \exp(-|j-l|/\xi_j)
\end{equation}
with localization length $\xi_j\approx 1/(2\log h)$.

When $J \ne 0$ one might expect the resulting interactions
to destroy the Anderson orbitals, either by hybridization or by an effective, self-induced dephasing,
resulting in a system that comes to (local) thermal equilibrium.
But in the limit of large disorder $h \gg J$
this model has been rigorously shown to exhibit many-body localization,
in which eigenstate correlations are short-ranged and dynamical correlations do not decay~\cite{imbrie_many-body_2016}.
For $h \sim J$ competition between localization and interaction-induced dephasing results in a phase transition from that many-body localized phase to one in which the system quickly comes to equilibrium~\cite{Basko_2006_MBL,Oganesyan_2007,Pal_2009}.
This phase is described by the eigenstate thermalization hypothesis \cite{Deutsch_1991_ETH,Srednicki_1994_ETH,Tasaki_1998_ETH,Rigol_2008_ETH,dalessio_quantum_2016}
which states that local operators have energy eigenstate expectation values very similar to their Gibbs state expectation values.

\subsection{Quantum Facilitated Network Model} \label{ss:fac-net}
We wish to fully take into account Anderson physics,
while selectively probing the effects of interactions and resonances.
We therefore work in the basis of the Anderson orbitals.
In this basis the fermion Hamiltonian \eqref{eq:Heisenberg} has the form%
\footnote{
This is the exact form of the Hamiltonian \eqref{eq:Heisenberg}. That Hamiltonian is quartic in the single-fermion operators $c^\dagger_j, c_j$;
rewriting it in terms of the Anderson-orbital fermion operators $f^\dagger_j, f_j$
modifies the spatial structure
(now encoded in the $U,V,W$)
but gives a Hamiltonian that is still quartic.
}
\begin{align}
    \label{eq:AndersonFull}
\begin{split}
    H &= \Hnet + \sum_{jklm} W_{jklm} c^\dagger_j c^\dagger_k c_l c_m \;,
\end{split}
\end{align}
where $\Hnet$ is the \textit{facilitated network Hamiltonian}
\begin{align}
    \Hnet = \sum_j \varepsilon_j n_j +\sum_{j,l}U_{jl} n_j n_l +\sum_{jkl} V^{(k)}_{jl} c^\dagger_j n_k c_l\;.\label{eq:NetworkHam}
\end{align}
This Hamiltonian enables hopping between different Anderson orbitals $j,l$ with rate $V^{(k)}_{jl}$ only if the hopping is facilitated by another particle in orbital $k$. In this sense, $\{V^{(k)}_{ij}\}$ describes a dynamical network in which a link between orbitals $i,j$ is active only if the orbital $k$ is occupied by a fermion. (See Fig.~\ref{fig:summary_figure}b for the average network topology).
It also includes a many-body correction to the energy of Anderson orbitals $\sim U_{jl}$.
The matrix elements $U_{jl}, V_{jl}^{(k)}$ are zero if two (or more) indices coincide.
The facilitated hopping term respects particle-hole symmetry as a result of a sum rule $\sum_k V_{jl}^{(k)}=0$.
For large disorder $h \gg 1$, the coefficients on the three and four-body terms fall off exponentially with the separation between the sites.
Additionally the average nearest-neighbor hopping terms fall off as $1/h$ and $1/h^2$ for three and four-body interactions, respectively~\cite{Laflorencie_2020}.
Deep in the MBL regime, $\varepsilon_j$ and $U_{j,l}$ accurately capture the local integrals of motion, with the l-bits converging to the Anderson orbitals.
This has previously motivated effective descriptions of the localized phase in which both three and four-body terms are neglected\cite{Laflorencie_2020,De_Tomasi_2019} or truncated to finite link-weight \cite{prelovsek_2018,prelovsek_2020}.

Given the stronger suppression of two-particle hopping terms $\sim c^\dagger_j c^\dagger_k c_l c_m$ compared to the facilitated single-particle hopping $\sim c^\dagger_j n_l c_m $  in the MBL phase, we consider a reduced model which discards two-particle hopping $W_{jklm}\rightarrow 0$, but retains the facilitated hoppings.
The resulting model describes a network of Anderson orbitals connected only by the facilitated hopping terms $V_{jl}^{(k)}$.
In this network model, all particle transport between Anderson orbitals requires the presence of a nearby occupied orbital to facilitate the hopping (Fig.~\ref{fig:summary_figure}a).

Figure~\ref{fig:summary_figure}b shows the emergent network structure, where the links represent effective hopping amplitudes $t_{jl}=\sum_k V_{jl}^{(k)} n_k$ which are time-averaged over several many-body configurations.
The color intensity of the arc between sites $j$ and $l$ is proportional to $\log |t_{jl}|$;
we use the log scale so all three disorder values can be plotted with the same scale.
The links become increasingly sparse as the disorder strength is increased, because increasing disorder reduces the spatial overlap of the orbitals.
It will also increase the difficulty of matching the resonance conditions.

We anticipate that neglecting the two-body hopping terms reduces particle hopping and therefore leads to stronger localization in the effective model compared to the full model. This would yield an enlarged localized phase in our facilitated hopping model compared to the full model. Nonetheless, for numerically accessible system sizes the apparent shift relative to the full model is negligible. We verify this by comparing the level statistics of the full Hamiltonian with the facilitated hopping model by means of exact diagonalization (see Fig.~\ref{fig:summary_figure}d). This suggests that the facilitated hopping model preserves the essential structure of the MBL transition and provides a more tractable, minimal model to study.

We further would like to anticipate that our network model describes a dynamical, state dependent network topology in real space. At a given time, the links depend on the instantaneous arrangement of particles, and change as the system evolves over time. This sets it apart from other network-inspired treatments, including so-called random regular graphs (RRGs)~\cite{Tikhonov_2016,Tikhonov_2019,Tikhonov_review,Garcia_Mata_2020}. RRGs typically are hypergraphs with fixed topology, and are used to model the entire many-body Hilbert space, i.e., each node corresponds to a basis element of Fock space. However, we can connect our approach to the RRG view of the MBL transition: we consider the hypergraph induced on Fock space in Sec.~\ref{ss:percolation}, and demonstrate that there is indeed a comparable crossover in the network structure of Fock space as we observe in the dynamics on the real-space network.

\subsection{Markovian Dephasing} \label{ss:bath}
In addition to the Hamiltonian dynamics, we consider an external source of dissipation, which gives rise to dephasing.
We desire dephasing that preserves the integrity of the Anderson orbitals.
This is achieved by adding random fluctuations to the energy of each individual Anderson orbital---that is, we add a fluctuating Hamiltonian
\begin{equation}
    \label{eq:Hamfluc}
    H_{\text{fluc}}(t)= \sum_l c^\dagger_l c_l \eta_{l,t}\;.
\end{equation}
We take the fluctuating energies $\eta_{l,t}$ to be Markovian, i.e., to have zero mean and to be uncorrelated for different times or positions:
\begin{align}
    \label{eq:noise-spec}
    \begin{split}
        \expctt{\eta_{l,t}} &= 0\\
        \expctt{\eta_{l,t} \eta_{l't'}} &= \tau^{-1}\delta_{ll'}\delta(t-t')
    \end{split}
\end{align}
where we write $\expctt{\cdot}$ for a noise average.
The full Hamiltonian is then the time-dependent $H(t)=\Hnet+ H_{\text{fluc}}(t)$.
The fluctuating bath has two effects: it dephases, i.e., destroys coherence between the Anderson orbitals,
and it temporarily brings into resonance orbitals that would otherwise have a small energy difference.
Both these effects combined mean that the dynamics of charge is well-described by a classical rate equation---the classical model of the next section.

From a Hilbert space perspective, the kinetic constraints imposed by the resonance condition fractionalize the Hilbert space into equal energy subspaces of configurations accessible by facilitated hoppings.
Loosening the resonance condition by allowing the system to borrow energy spoils this fractionalization.
Instead, the formerly disconnected subspaces become weakly connected on some characteristic timescale set by the bath.
One might then anticipate that the dynamics appear non-ergodic on this timescale, beyond which slow mixing via off-resonant transitions may thermalize the system.
We explore this crossover in both the real-time dynamics (Sec.~\ref{ss:dynamics}) and in the Hilbert space itself (Sec.~\ref{ss:percolation}).

\subsection{Classical Dynamical Network Model}\label{ss:classical_model}
Above, we have introduced an external bath that fluctuates on timescales fast compared with any in the system's Hamiltonian and which destroys coherence between successive hopping processes. The resulting dynamics can be well approximated by classical transitions with rate controlled by the hopping matrix elements $V_{ij}^{(k)}$.
To this end, we argue that charge dynamics in the model of Sec.~\ref{ss:fac-net}
is well-described by a related classical network model.

\subsubsection{Markov Transition Rates}\label{ss:justify-classical}

A detailed discussion of the effect of the bath is given in Appendix~\ref{app:off-resonance}.
Heuristically, one can understand it as follows.
Consider an Anderson-orbital particle density eigenstate
\[\ket{\bm k, i} = \prod_{l \in \bm k \cup \{i\} } c^\dagger_l \ket{0}\;.\]
Without the bath, one can estimate transition rates between $\ket{\bm k, i}$ and $\ket{\bm k, j}$ by Fermi's golden rule
\[ w^{(\text{coherent})}_{ij} = 2\pi |V_{ij}|^2 \delta(E_{\bm k,i} - E_{\bm k,j}) \]
where $E_{\bm k, j} = \sum_{\bm l \in \bm k \cup \{j\}} \varepsilon_l + \sum_{l,m \in \bm k \cup \{j\}} U_{lm}$ is the many-body energy of $\ket{\bm k, j}$
and
\begin{equation}
    V_{ij} = \sum_{k \in \bm k} V^{(k)}_{ij}
\end{equation}
is the total effective hopping.
The bath broadens the delta function to a Lorentzian with width $\tau$ by temporarily bringing orbitals into resonance;
the result is a rate
\begin{equation}
    \label{eq:hopping_rate}
    w_{ij} = 2 V_{ij}^2 \frac{\tau}{1 + \tau^2(E_{\bm k,i} - E_{\bm k,j})^2}\;.
\end{equation}

In Eq.~\eqref{eq:hopping_rate} we see that $\tau$ plays the role of a tuning parameter for the degree to which the resonance condition must be satisfied in order for particles to hop.
But $\tau$ has an additional, unwanted effect:
for exactly resonant transitions, for which $\Delta_{ij} = 0$,
the hopping timescale of Eq.~\eqref{eq:hopping_rate} goes as $1/\tau$.
We rescale all hopping rates by a factor of $\tau$, so that the timescale for resonant transitions is constant and $\tau$ acts only to set the scale for the resonance condition.
That is, we take the hopping rates to be
\begin{equation}
\label{eq:rescale-rates}
w_{ij} \rightarrow w_{ij}/\tau\;.
\end{equation}

This rescaling is familiar from studies of MBL systems coupled to physically local baths.\cite{Fischer_2016,Levi_2016}
Those studies found that the system's dynamics show good collapse
when one rescales time to (in our notation)
\[ \tilde t = t\tau^{-1}(V/h)^2 \]
By rescaling the rates as in Eq.~\eqref{eq:rescale-rates} we eliminate this straightforward dependence:
all the variation in $\tau$ we see goes beyond the physics of MBL systems coupled to local,
Anderson-orbital-destroying baths.

Let us briefly comment on the essential differences between the classical network model defined in Eq.~\eqref{eq:hopping_rate} and the resonant cluster renormalization group (RG) description of the MBL transition\cite{Vosk_2015,Potter_2015}.
Both approaches consider a classical effective model where resonant transitions play a central role, with \citet{Potter_2015} focusing specifically on the Anderson basis of a random-field spin-$\tfrac12$ system.
Starting from bare transition rates (analogous to $w_{jl}$), the RG scheme merges clusters which satisfy the resonance condition, accounting for line broadening as the scheme progresses.
Our scheme, on the other hand, emphasizes real time dynamics and so retains the configuration dependence of the energy differences $\Delta_{lm}$ stemming from the two-body interactions $U_{il}$.
This naturally imbues the network with a dynamical topology which is washed out in RG schemes.
Instead, the line broadening is encoded in both $\tau$ and the fluctuating interaction energy due to exploration of phase space by weakly localized particles.
This also is an essential difference between the network model and a corresponding mean-field rate equation where the local occupation may vary continuously between $0$ and $1$.

\subsubsection{Method: Kinetic Monte Carlo Dynamics} \label{ss:mc}

We have argued that the classical network model with transition rates Eq.~\eqref{eq:hopping_rate}
captures the dynamics of the quantum model Eq.~\eqref{eq:Heisenberg}
coupled to a peculiar bath.
We use a kinetic Monte Carlo algorithm to simulate the dynamics of this classical model.
We first prepare a random initial state in the Anderson basis
with occupation numbers $n_j \in \{0,1\}$ on each site.
We then implement the following update scheme:
\begin{enumerate}
    \item For all occupied sites, calculate the effective escape rate $\Gamma_i = \sum_j w_{ij}(1-n_j)$.
    \item Draw waiting times $t_i$ for each particle from an exponential distribution with scale $\Gamma_i$.
    \item For the fastest particle $i^* = \underset{i}{\mathrm{argmin}}\left( t_i\right)$, randomly select an allowed transition $i^*\rightarrow j$ with probability $\propto w_{i^*j}$.
\end{enumerate}
The physical time updates as $t \rightarrow t + t_{i^*}$.
\footnote{Code and and sample data analysis for this work is publicly available in a \href{https://github.com/kklocke/MBL_network}{GitHub repository}.}

For a fixed filling fraction $\bar n$ in system size $L$, the computational cost of the update step scales as $\mathcal{O}(L^2)$.
This can be achieved by storing the link weights $V_{ij}$ and updating after each particle transition.
The expectation value of the waiting time $t_{i^*}$ falls off as $1/L$, and so for a fixed physical time we anticipate that computational cost scales as $\mathcal{O}(L^3)$.
It is noteworthy, however, that for large system sizes the computational cost is dominated by the $\mathcal{O}(L^4)$ time required to calculate all entries of $V_{jl}^{(k)}$.
This could be reduced to $\mathcal{O}(L^3)$ by truncating the allowed range of such elements, but in the interest of keeping long-range hopping elements, we do not implement such a truncation.

\section{Results}\label{sec:results}

In this section we discuss the dynamics of particles in the classical network model, which we obtain from the kinetic Monte Carlo simulations of Sec.~\ref{ss:mc}.

We begin by describing the phase diagram as a function of disorder strength, the dephasing time scale, and the particle density.
We distinguish a frozen regime, in which a majority of the particles becomes immobile and localized in single Anderson orbitals, from an ergodic regime, in which the majority of particles is mobile and delocalized.
In Sec.~\ref{sec:Griffiths} we investigate and characterize rare ergodic bubbles, i.e. regions of mobile particles, which survive deep into the frozen regime.
We briefly comment on the relationship of these bubbles to those of the avalanche picture,
and the need for a timescale on which a system is considered frozen (localized).

In Sec.~\ref{sec:current_sec} we turn our attention to the nonequilibrium steady state (NESS) current induced by coupling the system to baths at the boundaries.
We observe a transition from diffusive to subdiffusive transport upon crossing over into the frozen phase.

\subsection{Dynamics and Localization in Single Orbital Observables}\label{ss:dynamics}

We aim to distinguish an ergodic regime, in which the particles are mobile and able to traverse the entire system, from a localized regime in which the particle dynamics freezes out. The former corresponds to a situation in which there is non-trivial overlap between nearby Anderson orbitals that are close to resonance, so particles can readily hop along the lattice. The latter then corresponds to the limit of large disorder ($h \gg 1$) and/or large dephasing time scale ($\tau \gg 1$), for which the system crosses over into a frozen regime where the timescales associated with particle hopping become arbitrarily long. We probe this localization-delocalization crossover through a number of observables, including the autocorrelation function of Anderson orbitals, the statistics of their occupation numbers and the statistics of waiting times between transitions.

Before delving into the details, we again emphasize that we see a crossover between two regimes, not a phase transition between two phases.
In particular, the location of the crossover depends on the timescales we consider.
Both of these observations are inherent to the model and to our focus on the occupations of individual Anderson orbitals, which are not exact eigenstates even in the localized regime:
for sufficiently (exponentially) long simulation times every particle will move eventually due to exponentially suppressed but non-zero transition rates $w_{ij}$.

\subsubsection{Autocorrelation Functions}

\begin{figure}
    \includegraphics[width=\columnwidth]{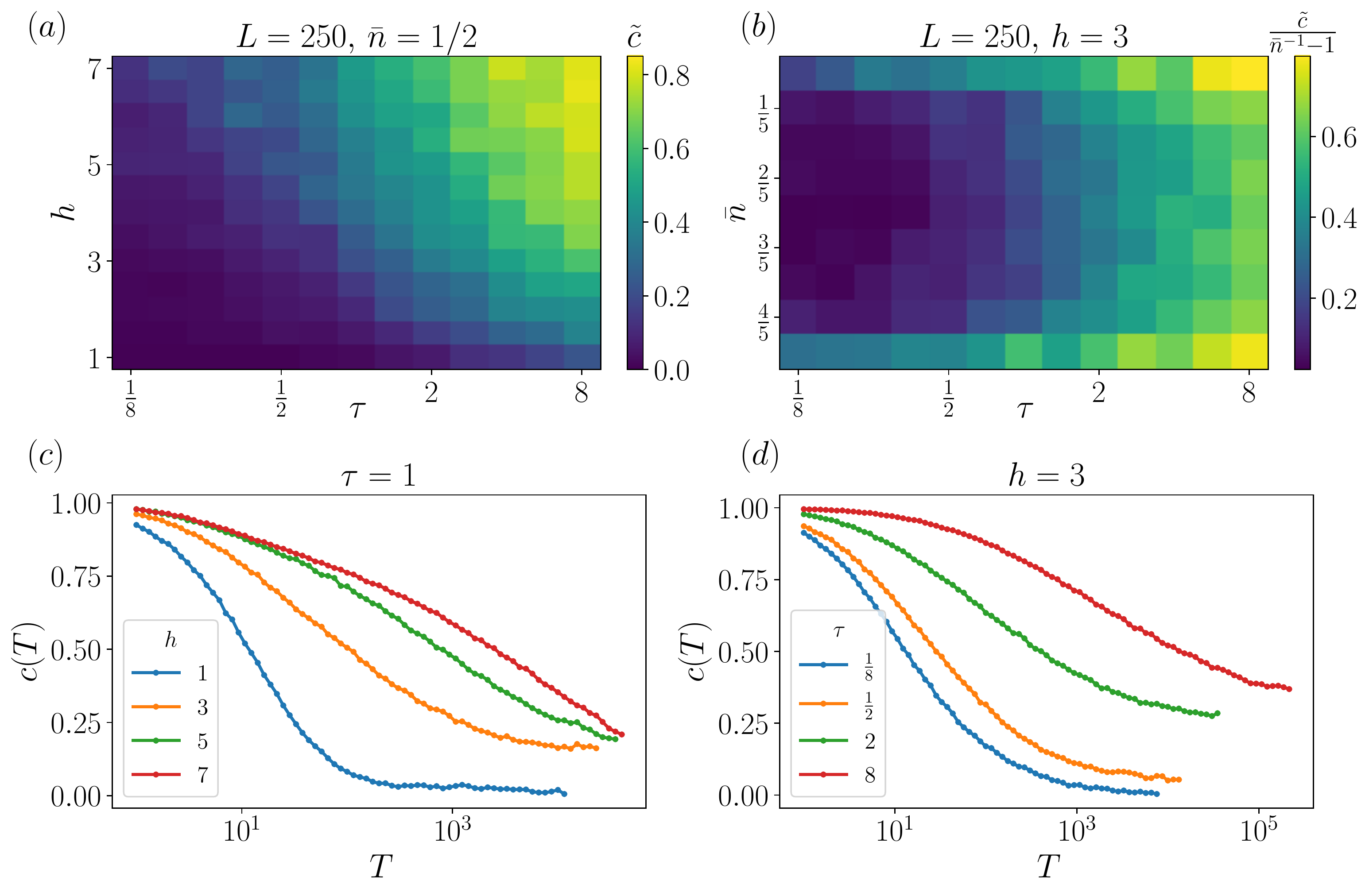}
    \caption{
    (a) Phase diagram on the $h{\rm -}\tau$ plane showing $\tilde c$ for a runtime $T \sim 10^4$ in the numerics.
    Correlations survive at late times only in a parameter regime where the disorder strength is large or the resonance condition is strictly enforced (i.e. large $\tau$).
    (b) Phase diagram in the $\bar n{\rm -}\tau$ plane for fixed disorder strength $h=3$.
    We show the renormalized correlation $\tilde c / ({\bar n}^{-1} - 1)$ such that perfect freezing has a value of $1$.
    At a fixed value of $\tau$ we see that the filling factor can tune between a frozen and active phase.
    (c,d) Temporal decay of correlations $c(T)$ as the disorder strength (c) or dephasing time (d) is varied.
    For strong disorder or small dephasing rates, the correlations survive to very late times.
    }
    \label{fig:correlation_phase_diagram}
\end{figure}
In the ergodic regime, one expects particles to rapidly and randomly move between different nearby Anderson orbitals.
Local correlations are thereby rapidly erased.
Conversely, in the frozen phase particles remain localized in their initial positions, giving rise to long-lived correlations.
The difference in behavior between the two regimes can be quantified by the autocorrelation function of the Anderson orbitals
\begin{equation}
    c(T) =\frac{1}{\bar n^2} \Big[\langle n_i(t)n_i(t+T) \rangle_{i,t} - \bar n^2\Big]\;.
\end{equation}
Here $\langle \cdots \rangle_{i,t}$ denotes the average over all sites $i$ in the chain and initial times $t$ and $\bar n=\frac{1}{L}\sum_i n_i$ is the average filling. For an individual orbital $i$, the autocorrelation function is  $\frac{1}{\bar n}-1$ ($=1$ for half-filling) for small $T$, and then starts to fluctuate at the characteristic rate at which particles hop into or out of this Anderson orbital. The characteristic correlation time is then defined as the time $T$ at which $n_i(t)$ and $n_i(t+T)$ start to become uncorrelated and for which $c(T)$ decays to zero (see Fig.~\ref{fig:correlation_phase_diagram}c).

The fraction of particles that remains frozen after some large time is then described by the integral
\begin{equation}
    \tilde c = \lim_{T\rightarrow\infty} \frac{1}{T} \int_0^T \dif T' c(T').
\end{equation}
Ergodic regions have a finite correlation time, such that their contribution to $\tilde c$ vanishes for sufficiently large times $T$.
Frozen regions, however, are characterized by a long correlation time, and therefore yield a non-zero contribution to $\tilde c$ for large times $T$. The frozen fraction $\tilde c$ therefore provides a diagnostic with which to map out the phase diagram: if $\tilde c=0$, the system is fully ergodic, while if $\tilde{c}=1$ it is completely frozen.

We first focus on a half-filled chain (filling fraction $\bar n=\tfrac12$) and map out the frozen fraction $\tilde c$ in the $h{\rm -}\tau$ plane in Fig.~\ref{fig:correlation_phase_diagram}a. The frozen fraction shows a crossover from the delocalized regime, with short-lived correlations and $\tilde c \approx 0$, to a frozen regime, where correlations may last arbitrarily long.
But even for parameters $h,\tau \gg 1$, the frozen fraction does not saturate to the upper bound $\tilde c \rightarrow 1$.
This corresponds to imperfect freezing, which we attribute to the presence of rare regions which remain mobile (delocalized) while the rest of the system's configuration is frozen.

Two related situations in the underlying quantum system can give rise to this imperfect freezing.
Both result from the formation of true many-body l-bits by interaction-induced hybridization of Anderson orbitals.
First, imagine that the number of orbitals involved is small.
In an Anderson basis, quantum dynamics due to these l-bits appears as precession---oscillation between Anderson orbitals.
That precession appears in our classical model as hopping between the few involved orbitals.
Second, imagine that a large number of orbitals is involved.
Then this hybridization may result in an ergodic grain
or a long-range resonant network.
Our classical model does not distinguish between these situations,
and it does not distinguish these situations from bath-induced transport between l-bits.

Additionally,
the mobile regions may facilitate hopping in nearby regions which would be otherwise frozen, effectively blurring the freezing transition into a smooth crossover. We observe that the mobile regions strongly modify the dynamics in the presence of a nonzero dephasing time; we will focus on the interplay between frozen and mobile regions below in Sec.~\ref{sec:Griffiths}.
This is consistent with the observation of persistent particle number fluctuations at large but finite disorder in the disordered Heisenberg  chain\cite{Sirker_2021a, Sirker_2021b, Sirker_2021c}.

If we move away from half-filling, we again observe a crossover between a mobile and a frozen region in the $\bar n{\rm -}\tau$ plane (see Fig.~\ref{fig:correlation_phase_diagram}b) for fixed random field $h$, which is reminiscent of a mobility edge.
The phase diagram is approximately symmetric about $\bar n=\frac12$ in Fig.~\ref{fig:correlation_phase_diagram}b.
(Residual asymmetry results from the fact that individual disorder realizations are not particle-hole symmetric.)

The crossover in $\bar n$ is unsurprising, since for sufficiently small filling fraction ($\bar n \ll \tfrac12$) the typical interparticle distance far exceeds the correlation length for the Anderson orbitals.
The associated hopping rates $w_{jl}$ are then exponentially suppressed, giving a very long timescale for transitions between different Anderson orbitals.
At $\bar n=1/L$ there is perfect freezing in the sense that the model prohibits hopping without a second occupied orbital to facilitate the process, i.e., one recovers single-particle Anderson localization.

If the system has only two particles there may be a subset of the configuration space where the particles remain active.
This is only possible when the inter-particle distance remains small so that they continuously facilitate transport in the same direction, or at the same position, giving short average times between particle hops.
While such states become rare in the thermodynamic limit (or for strong disorder), the system will be dynamically attracted towards configurations, which have a high mobility, and therefore continue frequent hopping.

\subsubsection{Distribution of the Time-Averaged Occupation}

\begin{figure}
    \centering
    \includegraphics[width=\columnwidth]{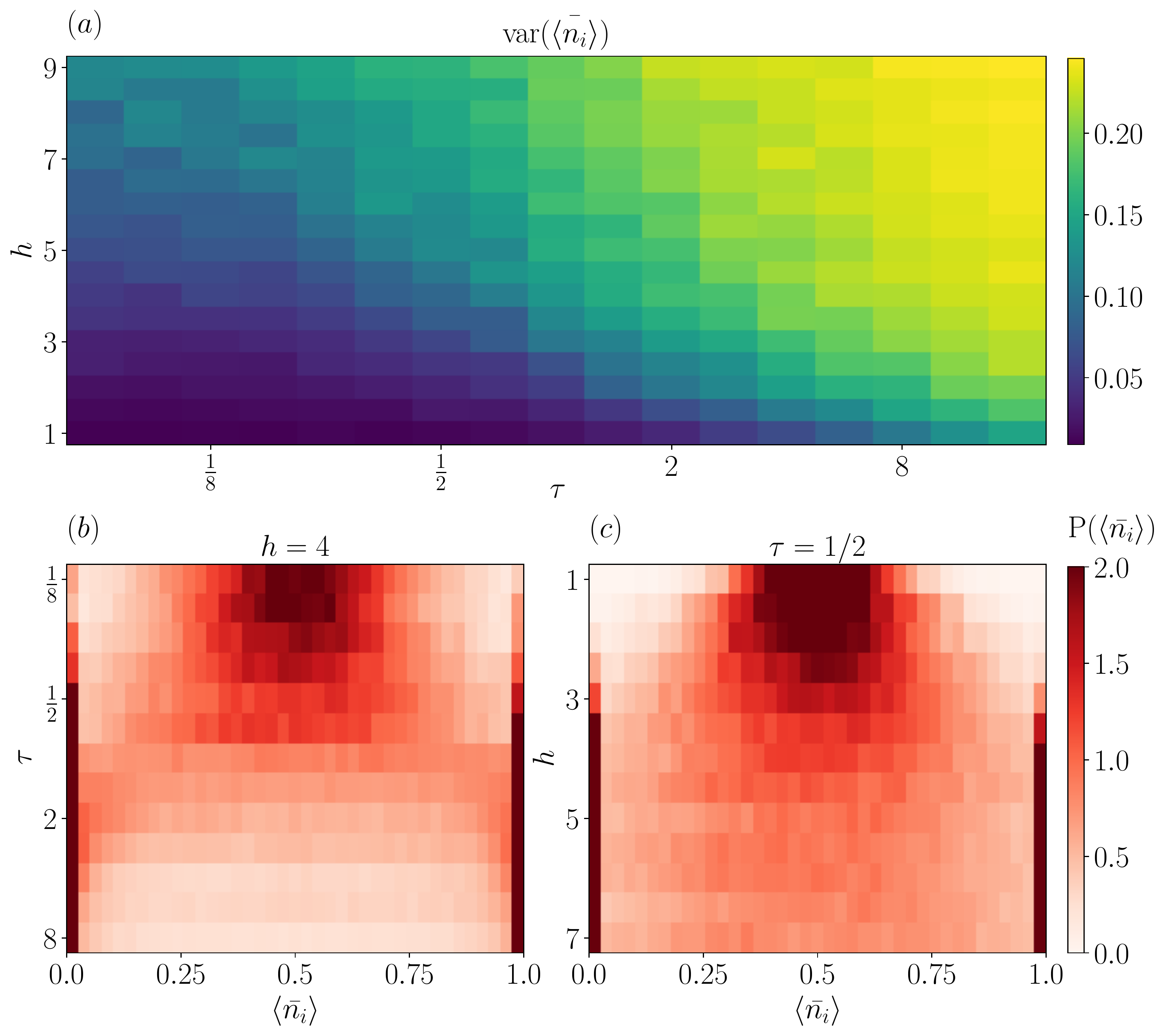}
    \caption{
    Time averaged site occupations $\langle \bar{n}_i \rangle$ for $N=250$ sites and a fixed time interval $T = 10^3$.
    (a) A phase diagram in the $h-\tau$ plane at half-filling showing the variance in the distribution of average occupations.
    At large $\tau$ or strong disorder $h$, the majority of sites are frozen and the variance approaches $0.25$.
    (b) The distribution of time-averaged site occupations as a function of $\tau$ for fixed disorder strength $h=4$.
    For small $\tau$, off-resonant transitions are permitted and the system remains ergodic over a range of disorder strengths, showing a unimodal distribution  about $\bar{\langle n_i \rangle} \approx \tfrac12$.
    At large $\tau$ the distribution becomes bimodal about $0$ and $1$, with remaining weight near $\tfrac12$ owing to rare-regions which remain active.
    (c) Similar results as in (b) but for fixed $\tau = \frac12$, showing the freezing transition at large disorder.
    It is noteworthy that for fixed $h$ in (b), at small $\tau$ there remain a number of sites which are frozen, even when most other sites are active.
    This should be contrasted with (c) where all sites are active at weak disorder.}
    \label{fig:occupations}
\end{figure}
In order to obtain a more detailed picture of the different dynamical regimes, we examine the distribution $P\left(\bar n_i \right)$ of the time-averaged orbital occupations,
\begin{equation}
    \bar{ n}_i = \lim_{T\rightarrow \infty}\frac{1}{T}\int_0^T \dif t\; n_i(t),
\end{equation}
where $T$ is once again the physical time interval.
We take filling fraction $\bar n = 1/2$.
The first moment of $P(\bar n_i)$ then is the filling fraction $\mathbb{E}[\bar{ n}_i]=\bar n=\tfrac{1}{2}$. The second moment, the variance
\begin{equation}\textrm{var}(\bar n_i) \equiv \mathbb{E}[\bar n_i^2] - \mathbb{E}[\bar n_i]^2,
\end{equation}
provides an alternative means to probe the active-frozen transition. For a region of mobile particles, we expect $\bar n_i=\frac{1}{2}$. For a completely frozen region, by contrast, we expect a bimodel distribution $\bar n_i=0,1$. The variance then is expected to be minimal, $\textrm{var}(\bar n_i)\rightarrow0$ in the ergodic regime, while it approaches its maximum value $\textrm{var}(\bar n_i)\rightarrow\tfrac{1}{4}$ for a frozen network.

We show in Fig.~\ref{fig:occupations}a that the variance of $P(\bar n_i)$ identifies a very similar phase structure to that obtained from the autocorrelation function in Fig.~\ref{fig:correlation_phase_diagram}a.
Once again one observes a crossover between an ergodic and a localized regime.
Only deep in the localized regime is the upper bound $\textrm{var}(\bar  n_i)\rightarrow \frac14$ reached.

A finer resolution of the dynamics in the crossover and in the frozen regime can be obtained by inspecting the distribution $P(\bar n_i)$ directly. It is displayed in Fig.~\ref{fig:occupations}b,c, and it shows a transition from an unimodal distribution, peaked at $\bar n=\tfrac{1}{2}$ in the ergodic regime, to a bimodal distribution in the frozen regime. But even for parameters that we expect to be deep in the frozen regime ($h,\tau \gg 1$), there remains a nonzero probability density for $\bar n_i \approx \tfrac 12$, which we attribute to the presence of rare ergodic regions.

Like the autocorrelation function, the averaged site occupations tend to overestimate the extent of the active phase: (i) even orbitals with rare activity may appear active, if they are populated/unpopulated for an equal amount of time, which would shift $\bar n_i\rightarrow \tfrac{1}{2}$, and (ii) rare, isolated regions with a small number of resonant Anderson orbitals will always appear with a nonzero probability. This yields a nonzero density of small and isolated clusters of mobile particles, which enter the statistics of $\bar n_i$ but will have no impact on the majority of frozen Anderson orbitals. In both scenarios, despite the fact that the vast majority of the evolution is frozen, the time-averaged site occupation would suggest that the system is not frozen.

\subsubsection{Waiting Times}

\begin{figure}
    \centering
    \includegraphics[width=\columnwidth]{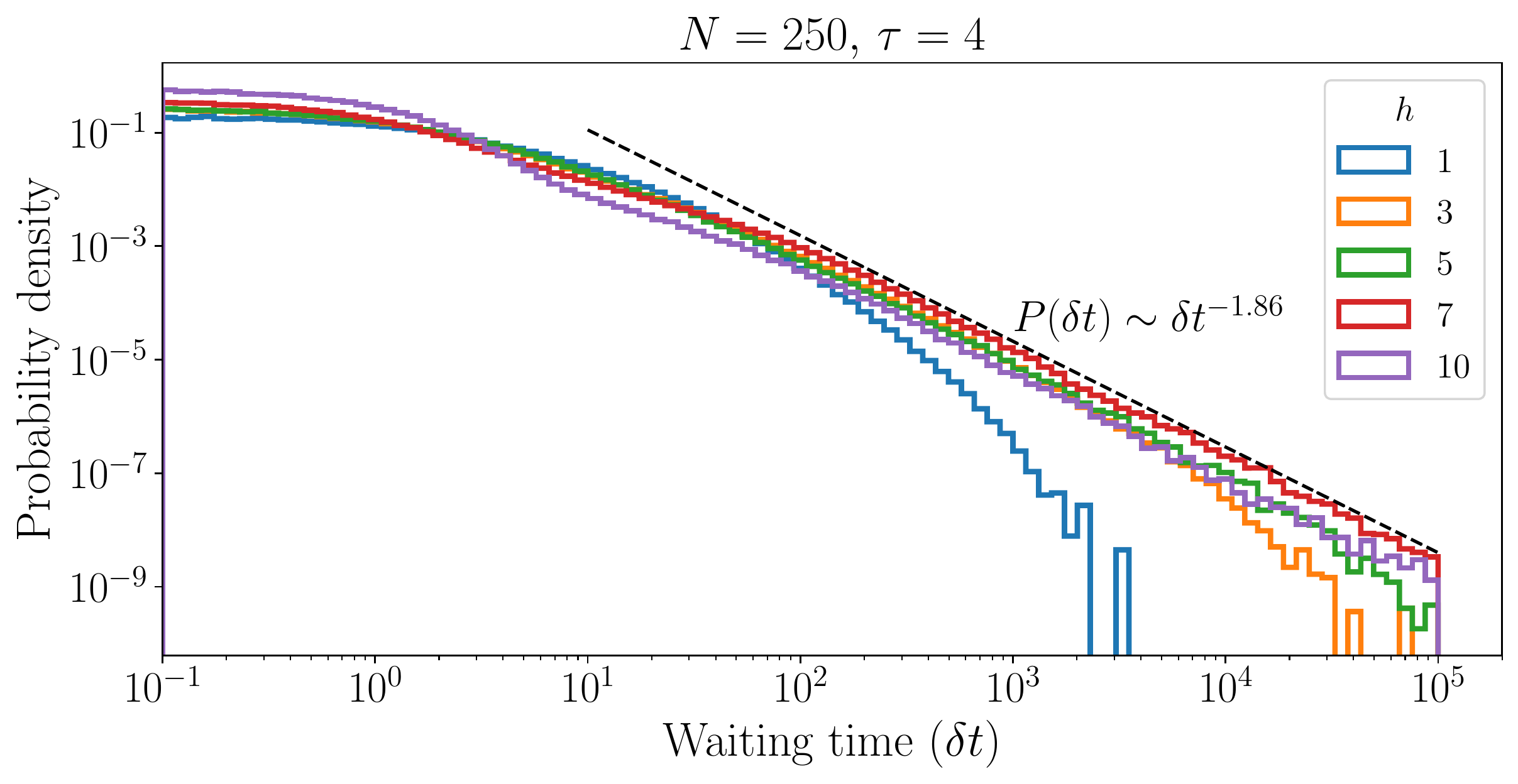}
    \caption{
    Distribution of (onsite) waiting times between particle hoppings for various disorder strengths with a fixed energy window $\tau = 4$.
    At weak disorder, the distribution falls off rapidly such that the typical waiting time is much shorter, as one would expect in an ergodic regime.
    The freezing transition at large disorder is accompanied by the appearance of an algebraic distribution of long waiting times, decaying with an exponent of approximately $-1.86$ (see dashed black line for visual guide).
    }
    \label{fig:waiting_times}
\end{figure}

To address the mentioned limitations of the time-averaged occupation as a probe,
we inspect the distribution of
{\it waiting times}---that is, the time $\delta t$ an individual particle stays in the same Anderson orbital before hopping to another orbital.
The distribution of waiting times is displayed in Fig.~\ref{fig:waiting_times}, for varying disorder strengths and fixed dephasing time $\tau=4$.
For all disorder strengths, there is appreciable weight at short waiting times ($\delta t \lesssim 1$).
We attribute this behavior again to the presence of small clusters, which can undergo frequent dynamics due to a resonance in the potential energy.

For large waiting times $\delta t \ge 1$, however, the distribution shows a significant dependence on the disorder strength.
For small disorder, the distribution is cut off at a finite time, indicating a lower bound for the transition rates in the ergodic phase.
Increasing the disorder strength, the cut off shifts to larger waiting times, and eventually reaches the total simulation time (i.e., diverges). In this limit, the distribution approaches a power law with exponent approximately equal to $-1.86$ for $h\ge5$. Above this disorder strength, the long-time tail of the distribution no longer changes, and we expect a divergence of the mean waiting time for larger disorder strengths and in the thermodynamic limit.
This is reminiscent of the dynamics in spin glasses, where upon reaching the freezing transition, the width of the relaxation time distribution typically diverges.
In resonant cluster renormalization group studies, the distribution of effective tunneling rates plays a comparable role to the waiting times here.
The power-law tail at large waiting times can yield a diverging mean waiting time, which has been previously associated with subdiffusive energy transport\cite{Potter_2015} at the onset of the MBL transition.
As shown in Sec.~\ref{sec:current_sec}, our model shows a corresponding subdiffusive regime upon increasing the disorder strength above values of $h\approx 3$.

\subsection{Griffiths Effects}\label{sec:Griffiths}

For a wide range of parameters, the dynamics of the network is characterized by the simultaneous presence of frozen and mobile regions, which prevents us from unambiguously defining a frozen or localized phase. Instead it gives rise to a smooth crossover from the ergodic to a more and more (but never completely) frozen regime. In the following, we shift the focus away from single-orbital observables and instead seek to characterize the mobile and frozen regions, which we term {\it bubbles}, by quantifying their size, lifetime, and associated energy fluctuations.

\subsubsection{Defining Ergodic and Frozen Bubbles}

We consider an ergodic bubble to be a region (not necessarily contiguous) in which the typical particle activity rate is large.
In order to quantify the extent and distribution of such bubbles, we define a graph $\mathcal{G}=(V,E)$ with nodes ($V$) representing the Anderson orbitals and edges ($E$) with weight $E_{ij}$ equal to the number of times a particle has hopped between sites $i$ and $j$.
If this is taken over a physical time interval $T$, then the activity rates are $A_{ij} = E_{ij} / T$.
We can now introduce an activity threshold $\epsilon$ and define a new edge set $\tilde E$ such that $\tilde E_{ij} = \Theta(A_{ij} - \epsilon)$.
The connected components of the resulting graph now represent regions of high particle activity and are designated ergodic bubbles.
The \textit{size} of such active bubbles is given by the number of nodes in each connected component (not the physical diameter of the region spanned by the component).
Intervening sites $i$ with $A_{ij} < \epsilon$ for all $j$ then comprise the (contiguous) frozen regions.

An ergodic bubble defined in this way represents the classical analogue of what has been considered previously as an ergodic subregion in the quantum mechanical MBL setting. The latter are characterized as a set of states, connected by resonant couplings. We observe that, similar to the phenomenology in the quantum model, ergodic bubbles, once formed, tend to grow spatially towards the most active configuration~\cite{Thiery_2018,Morningstar_2019}. This results from the construction of our model, which treats the transitions $|\bm{n}_l\rangle\rightarrow|\bm{n}_m\rangle$ and $|\bm{n}_l\rangle\leftarrow|\bm{n}_m\rangle$ on equal footing, i.e., attributes the same transition rate to both processes. Transitions with higher transition rates are generally more likely to occur, so the system naturally evolves towards more active configurations.

The instantaneous onsite activity rate ($\sum_j A_{ij}$) for a chain of $L=250$ sites is shown in Fig.~\ref{fig:bubble_sizes}a.
Several regions of activity are visible, and separated by inactive regions. We also readily observe regions where extended periods of inactivity are separated by periods of intense activity.
We also note that near site 175 in Fig.~\ref{fig:bubble_sizes}a there is an active region which is \textit{not} contiguous.
In this case, a single (or a few) frozen particles facilitate the hopping between adjacent sites without actively participating in the dynamics.
This is a nascent example of a so-called ``resonant backbone'', i.e. an extended configuration with resonant hopping processes that is interrupted by seemingly inactive sites, which are, however, crucial for the facilitation of nearby activity.

Before analyzing the distribution of bubble sizes, we would like to mention that previous works have suggested both that at criticality the distribution of ergodic bubble sizes should either follow a power-law~\cite{Thiery_2018} or exponential~\cite{Khemani_2017} distribution. The latter has been associated with the presence of an underlying resonant backbone~\cite{Khemani_2017}.

\subsubsection{Bubble Sizes}

\begin{figure}
    \centering
    \includegraphics[width=\columnwidth]{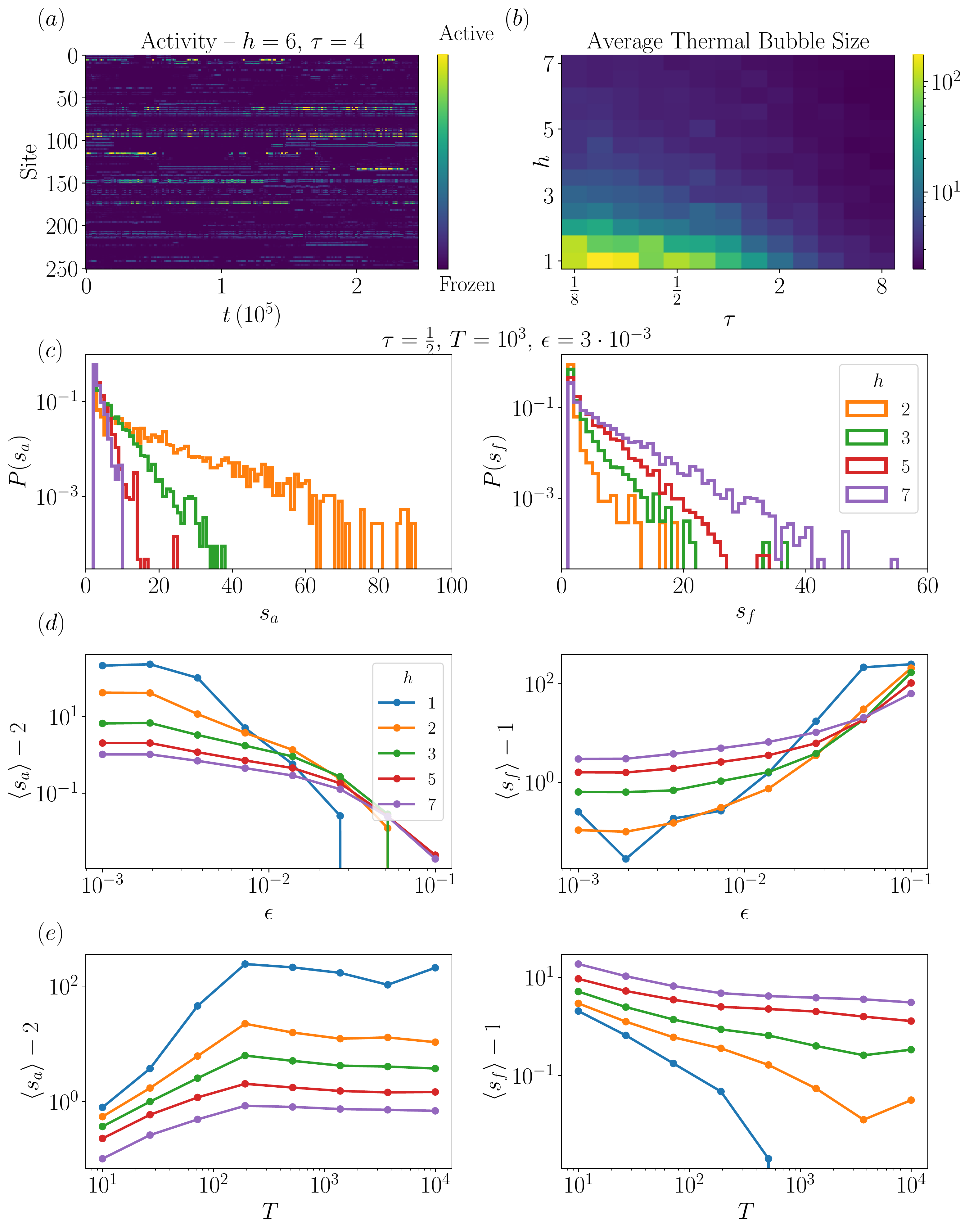}
    \caption{
    (a) Local activity rates during the evolution with $h=6$ and $\tau = 4$. Active versus frozen regions are readily distinguished.
    (b) Average active bubble size as a function of disorder strength and dephasing time.
    Here bubbles are determined by considering the activity graph over fixed time intervals of $T=10^3$ with a threshold $\epsilon = 3\cdot 10^{-3}$.
    There is a clear crossover from completely delocalized particles to rare thermal bubbles.
    (c) Distribution of bubble sizes as determined by the activity rate during a fixed physical time interval $T=10^3$ with cutoff $\epsilon=3\cdot 10^{-3}$ for various disorder strengths.
    (d,e)
    Mean frozen/active region size ($s_{f/a}$) as a function of (d) the activity rate threshold for qualifying as an active bubble or (e) the time interval over which activity rate is computed.
    The minimum possible bubble size has been subtracted off to highlight the scaling.
    Poor averaging for large $T$ at small disorder strength is responsible for deviations from the trend in $\langle s_f \rangle$ in (e).
    }
    \label{fig:bubble_sizes}
\end{figure}

Following the procedure above, we extract the size of frozen and active regions from the dynamics.
In the active regime, the whole system is ergodic, giving an extensive mean ergodic bubble size.
Crossing over into the frozen regime, thermal regions become rare and small, tending toward an average size of order unity (Fig.~\ref{fig:bubble_sizes}b).
Around the extended localization-delocalization cross-over, the active and frozen bubble sizes are both exponentially distributed (see Fig.~\ref{fig:bubble_sizes}c).
The distribution function $P(s_{\text{a,f}})$ for the size, i.e.
number of active sites,
of the active bubbles  ($s_{\text{a}}$) or the size of the frozen bubbles ($s_{\text{f}}$) is then of the form
\begin{equation}\label{eq:bubble_dist}
    P(s_{\text{a,f}})=A_{\text{a,f}}\exp(-s_{\text{a,f}}/v_{\text{a,f}}),
\end{equation}
where $v_{\text{a,f}}>0$ is the characteristic size for active (frozen) bubbles, which is subextensive in the crossover regime.

This observation is fairly robust with respect to the finite timescale of the simulation (see Fig.~\ref{fig:bubble_sizes}e).
However, if we compute the average bubble size
\begin{equation}\label{eq:avg_bub_size}
    \langle s_{\alpha}\rangle=\int \dif s_\alpha s_\alpha P(s_\alpha)
\end{equation}
from the numerical simulations, we find that it remains sensitive to the choice of threshold $\epsilon$ (Fig.~\ref{fig:bubble_sizes}d).
For sufficiently large (small) $\epsilon$, all regions are considered frozen (active).
However, over a broad range of $\epsilon$ there is an approximate power-law relationship between $\epsilon$ and the mean bubble size.
This implies the same power-law relationship between the {\it (in-) activity time scale }(threshold) $\epsilon^{-1}$, and the characteristic length scale $v_{\text{a}}$ of bubbles with precisely this activity time scale.
In the corresponding regime, this yields a dynamical scaling relation for active regions $v_{\text{a}}\sim \epsilon^{-\nu_{\text{a}}}$.
From fitting the numerical results, we find that the exponent $\nu_{\text{a}}$ decreases monotonically with increasing disorder strength, i.e. for $\tau=\frac12$ it starts close to $\nu=2.5$ for weak disorder ($h=1$) and continuously decreases for stronger disorder, taking values $\nu_{\text{a}} = 1.0$ ($\nu_{\text{a}} = 0.18$) at $h=2$ ($h=5$).

The results in Fig.~\ref{fig:bubble_sizes}c-e are averaged over several disorder realizations and initial particle configurations.
When the typical frozen bubble size $\langle s_f \rangle$ and number of frozen bubbles becomes small, individual realizations of the system are more susceptible to large fluctuations.
In particular, an exponentially rare large frozen region may appear and act as a bottleneck in the system.
Such rare fluctuations are responsible for the non-monotonic behavior at $h=2$ in Fig.~\ref{fig:bubble_sizes}e.
With additional disorder averaging or in the thermodynamic limit, we anticipate that this will approach a smooth curve with respect to varying $T$.

\subsection{Nonequilibrium Steady State Currents}\label{sec:current_sec}

In order to characterize the transport behavior of our network model, we turn our attention to the dynamics in the presence of a particle source and sink.
We therefore introduce a non-zero tunneling probability, with which particles are tunneling into the network at one end (particle source) and tunneling out of the network at the opposing end (particle drain).
We fix the tunneling rates to be equal on both ends so that particle-hole symmetry is maintained on average (i.e. the time averaged particle density remains near $\bar n\approx\tfrac12$).
Under these conditions we measure the nonequilibrium steady state (NESS) particle current, which is given by
\begin{equation}
    j(h,\tau,L) = \lim_{T\rightarrow \infty} \frac{N_\textrm{lost}(T)}{T}.
\end{equation}
Here $N_\textrm{lost}(T)$ is the number of particles tunneling out of the system in a time interval $T$.

\begin{figure}
    \centering
    \includegraphics[width=\columnwidth]{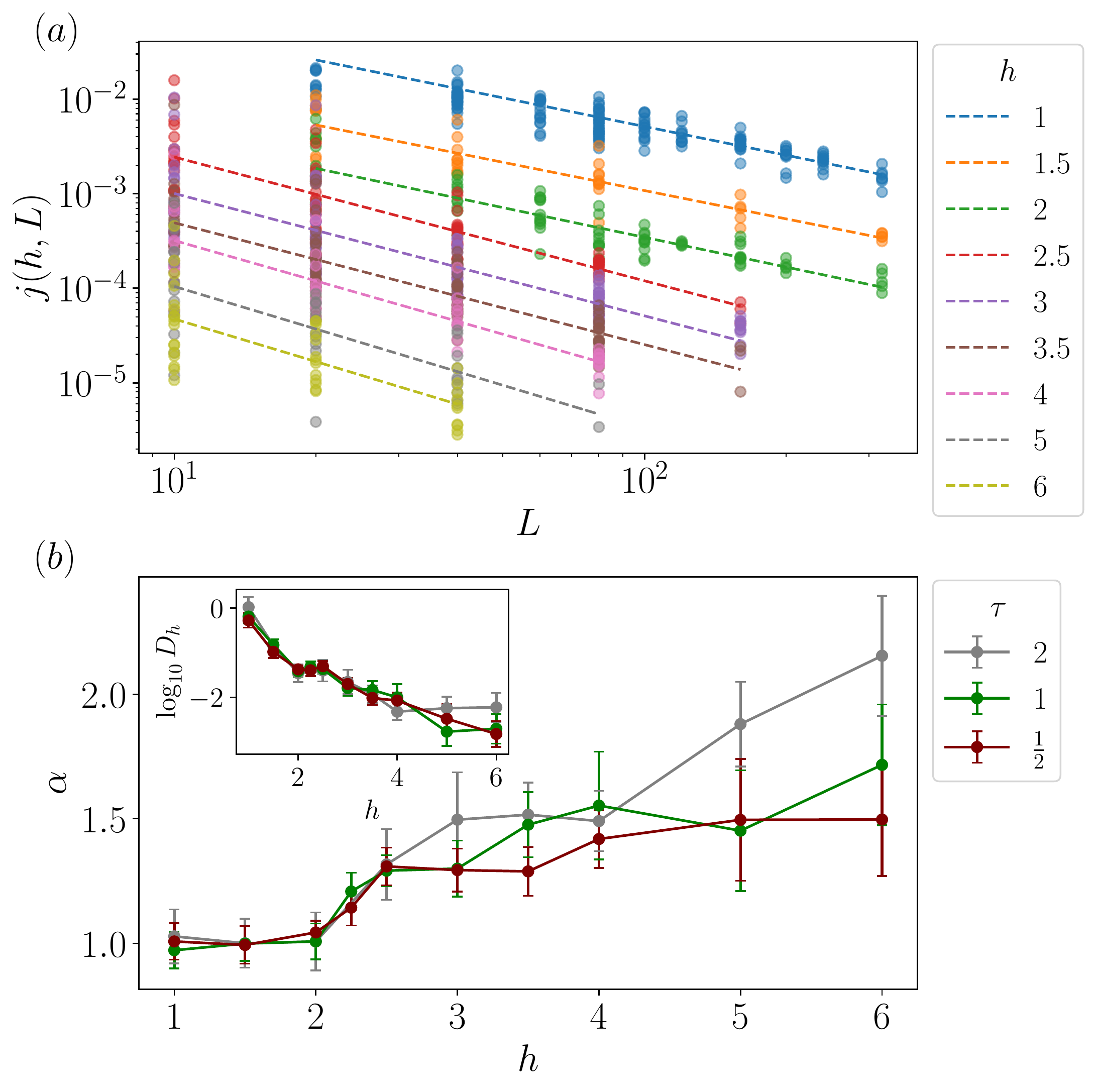}
    \caption{
    Scaling of the particle current with disorder strength ($h$) for a fixed energy window $\tau = \tfrac12$.
    Faint circles correspond to individual data points from different Monte Carlo trajectories.
    (a)
    Scaling of the nonequilibrium steady state current $j$ with system size as a function of the disorder strength for fixed $\tau = \tfrac12$.
    The data show a power-law scaling $j(h,L) \sim D_h L^{-\alpha}$, with $D_h$ a disorder dependent diffusion coefficient and $\alpha$ the scaling exponent.
    Several initial configurations and disorder realizations are taken for each disorder strength $h$ and system size $L$.
    (b)
    The fitted scaling exponent $\alpha$ shows a transition from diffusive ($\alpha \approx 1$) to subdiffusive ($\alpha > 1$) for disorder strength exceeding $h \approx 2$ for $\tau=\frac12$.
    (inset)
    The diffusion coefficient $D_h$ decays exponentially with disorder strength in the ergodic regime ($h \leq 2$) and the frozen regime ($h > 2$), albeit with different scales.
    Deeper into the frozen regime it becomes increasingly computationally intensive to obtain accurate estimates of the scaling exponent as the necessary runtime rapidly increases with $L$ and $h$.
    }
    \label{fig:noneq_current}
\end{figure}

This current depends sensitively on both the disorder strength $h$ and the dephasing time $\tau$.
For a fixed dephasing time $\tau$ we find that the NESS current $j(h,L)$ is well described by a scaling function $j(h,L) \sim D_h L^{-\alpha}$ for system sizes $L$ much larger than the correlation length.
Here $D_h$ is a disorder dependent diffusion coefficient which falls off rapidly with stronger disorder, and $\alpha$ is the scaling exponent with respect to system size.
As seen in Fig.~\ref{fig:noneq_current}, the exponent is fixed at $\alpha \approx 1$ for weak disorder, corresponding to diffusive transport.
At larger disorder, around $h=2.5$, the transport becomes subdiffusive, with $\alpha > 1$.
From the $\tau = \tfrac12$ data in Fig.~\ref{fig:noneq_current}b we see that the scaling of the NESS current shows a clearer transition than the crossover seen in the correlation and variance in occupation.
Upon varying $\tau$ the critical disorder strength at which we observe the onset of subdiffusion varies in a manner similar to the crossover observed in local observables (see Fig.~\ref{fig:correlation_phase_diagram}).

In the full quantum model, subdiffusive transport has been observed in the vicinity of the MBL transition\cite{Thiery_2018,Potter_2015,Vosk_2015,znidaric_diffusive_2016,Agarwal_subdiffusion,Bar_subdiffusion}
and may be attributed to Griffiths effects.
Despite not showing a sharp transition between the active and frozen regimes, our data do reproduce this crossover in transport properties.
Much as there is no perfect freezing observed in the correlation (Fig.~\ref{fig:correlation_phase_diagram}), transport does not completely vanish in the frozen phase.
In the parameter regime explored here we observe subdiffusion everywhere in the frozen phase, consistent with Ref.~\onlinecite{Taylor_2021}, where onsite dephasing in the disordered XX model yields a diffusion-subdiffusion transition.
Other works report that non-zero dephasing in the local, lattice basis may eliminate subdiffusive transport in MBL systems, and instead gives rise diffusive transport for all disorder strengths\cite{Znidaric_2010,znidaric_dephasing_2017}.
By tuning the dephasing $\tau$ in the classical network model, we are able to better probe the role of off-resonant transitions in facilitating this transition in particle transport.

\section{Local Energy Fluctuations and Self-Thermalization}

So far we have treated the dephasing time $\tau$ as an external parameter, resulting from an external bath,
and we have used it as tunable parameter with spatially uniform effect. However, if we consider a closed quantum system, a similar type of dephasing effect is caused by rapidly fluctuating particle densities in thermal bubbles. Orbitals inside and in close vicinity to a thermal bubble experience fast temporal fluctuations of the local interaction energy $E_{{\bf k},i}$, which acts similar as the randomly fluctuating onsite energy in Eq.~\eqref{eq:Hamfluc}. Due to the inhomogeneous distribution of the thermal bubbles, this will result in dephasing, which we can characterize by a spatially varying local dephasing time $\tau(x)$.
This time is then interpreted as the dephasing of many-body hopping matrix elements (in the interaction picture)
$V_{lm}(\{n_i\})e^{i(E_{{\bf k},l}-E_{{\bf k},m})t}$
between states $l,m$ due to rapid fluctuations of the occupations on nearby sites ${\bf k}$, i.e., the many-body energy difference $ E_{{\bf k},l}-E_{{\bf k},m}$ for the transition $l\leftrightarrow m$ undergoes temporal fluctuations due to changes in the configuration ${\bf k}$.
If these changes happen on a faster scale than the typical hopping time, the fluctuations have the same effect as the dephasing bath, c.f. Eq.~\eqref{eq:Hamfluc}.
In order to compare such a {\it self-generated} dephasing time with our external dephasing time $\tau$, we analyze in this part the dynamical energy fluctuations in active and in frozen regions. These fluctuations are conceptually the first step in an iterative solution of a self-consistent mean field theory like that of Ref.~\onlinecite{gopalakrishnan_mean-field_2014}.

We start by defining an instantaneous, ``local'' energy for a region $\ell$ via
\begin{equation}
    E_\ell =\bra{\bm n}H|_{\ell} \ket{\bm n}\;.
\end{equation}
Here the local energy $H|_{\ell}$ contains
(i) all single-particle terms of the facilitated network Hamiltonian (Eq.~\eqref{eq:NetworkHam}) in the subregion $\ell$ and
(ii) all many-body operators of Eq.~\eqref{eq:NetworkHam} that contain at least one density operator $n_l$ with $l\in \ell$.
From the local energy $E_\ell$, we determine the density of dynamical energy fluctuations from
\begin{equation}
    f_\ell^2 = \frac{\overline{ E_\ell^2} - \overline{E_\ell}^2}{s_\ell^2}.
\end{equation}
Here the overbar indicates time-averaging and $s_\ell$ is the number of sites contained in the subregion (active or inactive).
It is worth noting that since subregion $\ell$ need not be contiguous (for thermal bubbles), then $E_\ell$ and $f_\ell$ are not wholly local quantities.
However, this should only be relevant when the disorder is sufficiently weak that long-range resonant hoppings are possible.

\begin{figure}
    \centering
    \includegraphics[width=\columnwidth]{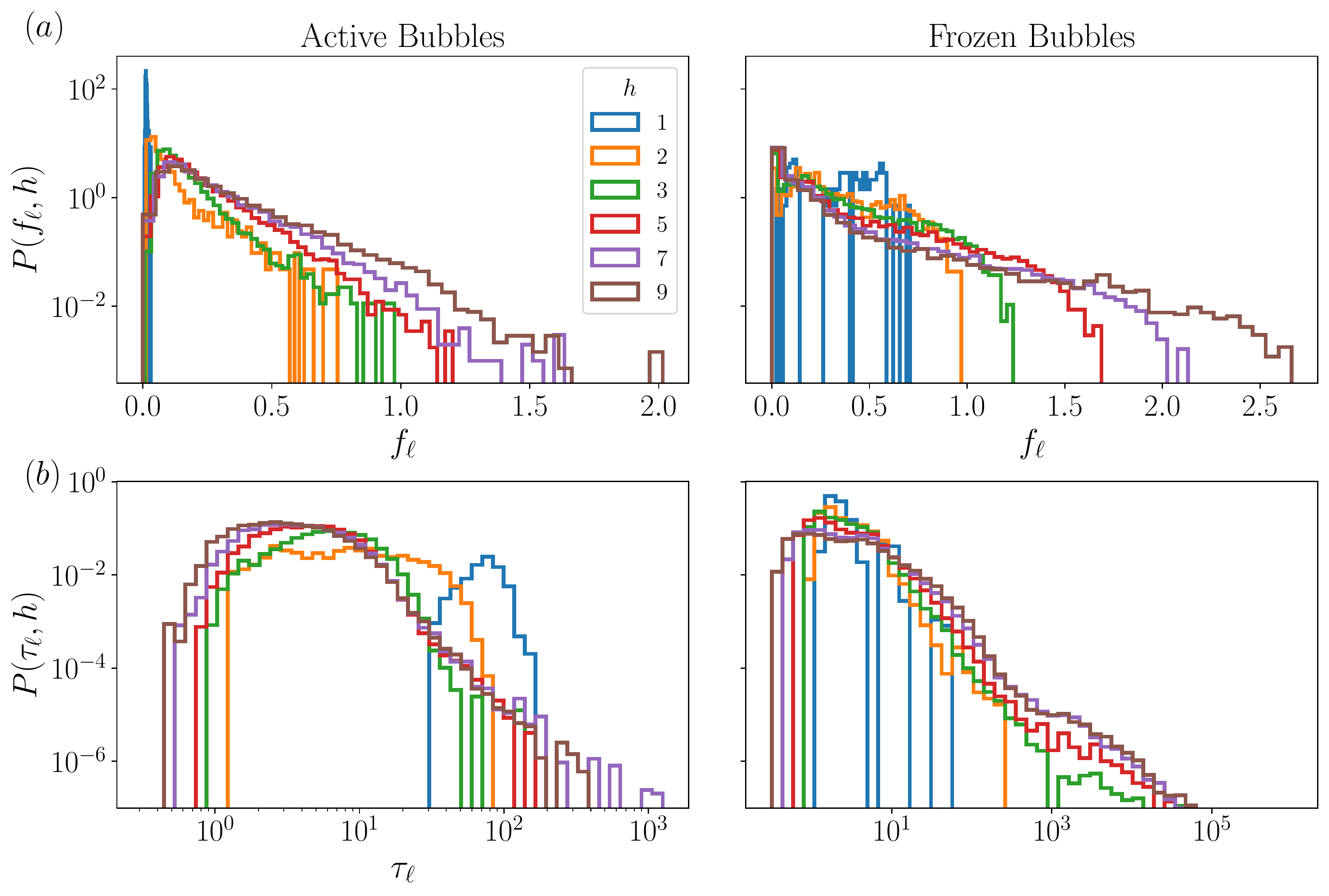}
    \caption{
    (a) The distribution of energy density fluctuations $f_\ell$ in active (left) versus frozen (right) regions for $L=250$ and $\tau = \tfrac12$ calculated over time intervals $T=10^3$.
    (b) The distribution of effective dephasing times $\tau_\ell \approx 1 / f_\ell$ in active and frozen regions.
    We see that frozen regions show an effective dephasing time which may be several orders of magnitude larger than that seen in the active regions.
    }
    \label{fig:energy_fluctuations}
\end{figure}

As shown in Fig.~\ref{fig:energy_fluctuations}, at weak disorder there are finite energy fluctuations in both active and frozen regions.
At larger disorder, however, frozen regions have an appreciable probability of vanishingly small fluctuations.
Nonetheless, there remains a finite probability for appreciable energy fluctuations in the frozen regions, with the distribution $P(f_\ell, h)$ falling off exponentially with increasing $f_\ell$.
This should be understood as arising from rare off-resonant transitions occurring within these regions.
For frozen regions, these rare hoppings necessarily are accompanied by a large energy change.
If this were not the case, then the region would experience far more particle transitions and become thermal.
The narrower distribution of fluctuations in the active regions then can be understood as arising from the frequent particle transitions between nearly resonant sites, which incurs only a small energy difference.
Unlike in frozen regions, however, at increasing disorder the probability weight increases in the tail of the distribution.

Here we observe an important qualitative distinction between active and inactive bubbles: (i) for active regions $\ell$, the energy fluctuations are due to genuine exploration of the configuration space \textit{within} the subregion $\ell$, while (ii) for an inactive region $\ell$ a significant fraction of energy fluctuations has to be attributed to the configuration changes \textit{outside} the region $\ell$, i.e., to fluctuations in nearby active regions.
In particular, let $\mathcal{P}_\ell$ be the projector onto the region $\ell$.
If we restrict the computation of the local energy $E_\ell$ to operators which are acting {\it only} on the subregion $\ell$ via $E_\ell=\bra{\bm n}\mathcal{P}_\ell^\dagger H \mathcal{P}_\ell \ket{\bm n}$, the density fluctuations remain nearly unaffected in active regions but are significantly reduced in the inactive regions (by about 25\%).

Active sub-regions $\ell$ thus lead to fast variations of the phase of the hopping matrix elements $\sim e^{i(E_{{\bf k},l}-E_{{\bf k},m})t}$ and generate an effective dephasing time $\tau_\ell\approx1/f_\ell$ for the active regions themselves and also for nearby inactive regions. The latter mechanism allows otherwise frozen regions to experience a broader spectrum of energy levels (smeared spectral lines), giving a greater likelihood of satisfying a resonance condition and having a particle hop.
This can be viewed as a classical, dynamical analogue to the renormalization schemes developed to describe how thermal inclusions grow and destabilize an otherwise nonergodic phase in isolated systems\cite{Zhang_2016,Vosk_2015,Potter_2015,Thiery_2018}.

We anticipate that in a closed system, where all dephasing times $\tau_\ell$ vary spatially and are self-generated (mainly by active regions), the distinction between frozen and active regions would become more sharp and the presence of entirely frozen regions in the chain would be more likely. Here, where we work with a global, externally determined dephasing time $\tau$, it acts as a threshold which suppresses complete localization.
Whether or not a sharp freezing (or localization) transition would occur for a closed system with self-generated dephasing times remains to be investigated and we emphasize that it is closely related to conditions on stability of a nonergodic phase with thermal inclusions.

\section{Weak Fractionalization in Fock Space}\label{ss:percolation}

Thus far we have focused on the (classical) real-time dynamics on a facilitated Anderson network, with a dynamical connectivity that depends on the real-space occupations. The original, quantum mechanical MBL transition, however, is mostly seen as involving the entire Fock space hypergraph, whose dimension grows exponentially with the system size.
Recent works~\cite{prelovsek_2018,prelovsek_2020,Roy_2019a,Roy_2019b} have argued that imposing a resonance condition to truncate the set of allowed particle hoppings
leads to a percolation transition in the Fock space hypergraph, which can then be viewed as a proxy for the MBL transition. In this framework, the real-space dynamics on the facilitated Anderson network can be seen as a random walk on the projection of the hypergraph in Fock space onto a real-space network in the Anderson basis.
Whereas the topology of the facilitated Anderson network is time-dependent (changing as particles hop), the hypergraph in Fock space is static and depends only on a particular disorder realization $\{h_j\}$. We will show that both pictures, the freezing of the real-time dynamics in a classical facilitated Anderson network and the percolation transition in Fock space of $\Hnet$ from Eq.~\eqref{eq:NetworkHam} are equivalent under appropriate conditions.

To this end, we construct a hypergraph $\mathcal{G} = (V,E)$ representing $\Hnet$ in Fock space.
Each Fock state then corresponds to a vertex $v \in V$. We introduce a cutoff activity rate $\epsilon$ and add the edge $(v_i, v_l)$ to the set $E$ if a transition matrix element $w_{il}$ (defined in Eq.~\eqref{eq:hopping_rate}) connecting the two Fock states $v_i$ and $v_l$ exceeds the threshold $w_{il}>\epsilon$. Depending on the parameters $h,\tau,\epsilon$, dropping edges with $w_{il}<\epsilon$ may cause the hypergraph to either remain in one giant connected component or to fractionalize into a finite set of mutually disconnected clusters.

This approach of constructing an effective hypergraph resembles previous ideas~\cite{prelovsek_2020,prelovsek_2018,Laflorencie_2020}, where a resonance condition for direct transitions $\left|\frac{V_{i,l}^{(k)}}{\Delta_{i,l}}\right|\ge\tilde\epsilon$ was used in order to determine whether an edge $(v_i,v_l)$ is added to the hypergraph or not. Despite this distinction, the clustering properties of the graph constructions are closely related. In our case, the threshold $\epsilon$ has a direct physical meaning. Like the determination of the ergodic bubble sizes, the inverse rate $\epsilon^{-1}$ sets a time-scale, below which we consider an edge to be present (active), and above which we consider it to be inactive (frozen).

\begin{figure}
	\centering
    \includegraphics[width=\columnwidth]{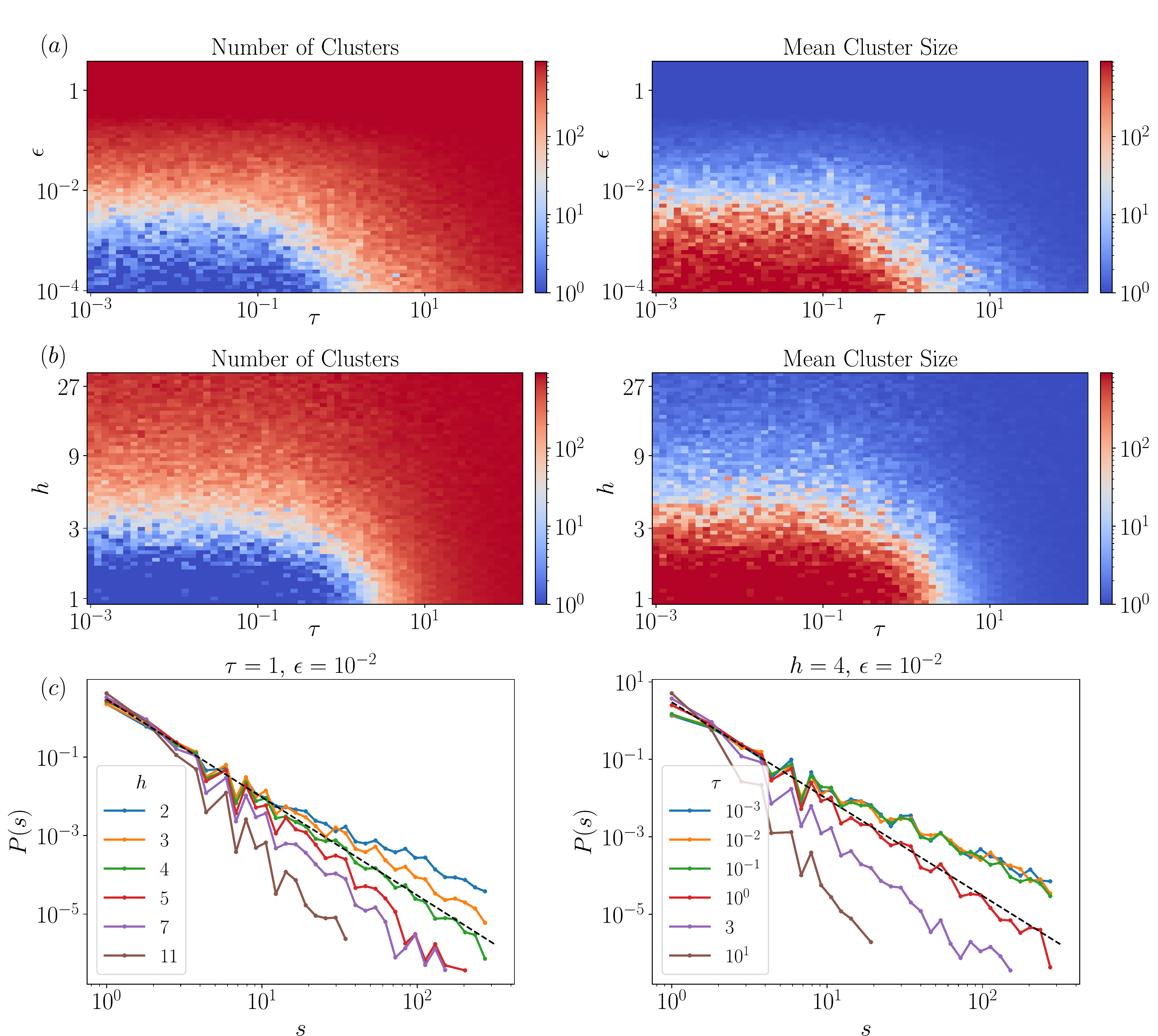}
    \caption{
	The mean number of clusters (left column) and cluster sizes (right column) in the hypergraph defined over Fock space at half-filling for $L=12$ sites.
	In (a) we vary the dephasing time $\tau$ and the edge-weight threshold $\epsilon$ with fixed $h=6$, whereas in (b) we vary $\tau$ and $h$ with fixed $\epsilon = 10^{-2}$.
	Both metrics show a phase boundary separating a regime where the hypergraph is fully connected (ergodic) from a regime where the hypergraph is broken into many disconnected components (non-ergodic).
	(c) The distribution of cluster sizes for varying disorder strength (left) and varying dephasing time (right) for system size $L=14$ averaged over several hundred disorder realizations.
	As we pass through the phase boundary the distribution falls off as a power-law with exponent $\approx 2.5$ (dotted black line).
	This is consistent with a percolation type transition, for which we expect a mean-field critical exponent of $5/2$ describing the distribution of cluster sizes.
    }
	\label{fig:fock_space}
\end{figure}

The hypergraph undergoes a percolation transition (or crossover) (Fig.~\ref{fig:fock_space}) when its giant connected component (marking an ergodic dynamics) decays into a large set of mutually disconnected clusters. Varying $\tau$ determines the strength of the energy conservation condition, and may split connected components of the hypergraph further into subgraphs of (approximately) equal energy. This induces a non-trivial resonant backbone structure, which is seen more naturally on the hypergraph (which collects clusters of connected Fock states) than on the real-space network. We find that the $h$---$\tau$ phase diagram obtained from the Monte Carlo dynamics is well-reproduced from the connectivity of the hypergraph (see Fig.~\ref{fig:fock_space}), with $\epsilon$ determining the position of the phase boundary.

Near the phase boundary, the distribution of cluster sizes $P(s)$ obeys a power-law $P(s)\sim s^{-\eta}$.
At the phase boundary the exponent is consistent with the (mean-field) exponent for percolation $\eta = 5/2$ (see dashed black line in Fig~\ref{fig:fock_space}c), in line with Ref.~\onlinecite{prelovsek_2020}.
Upon varying any of the parameters $\epsilon, \tau, h$, we generally move away from the critical point and see power-law scaling of cluster sizes with an exponent no longer equal to the value for percolation.
Nonetheless, for a fixed disorder strength $h$ near the critical value there is an extended range of $\tau$ for which the exponent remains stable near $5/2$.
To this end we denote this transition as being percolation-like, appealing to the picture in which it separates a regime with a giant connected component from a phase of disconnected clusters.

By examining the transition in Fock space we may better connect the dynamical freezing transition observed in our classical network model to the genuine MBL transition observed in the random field Heisenberg model and the facilitated hopping model.

\section{Discussion}

We started with the random-field Heisenberg model, which hosts a transition from an ergodic (ETH) phase to a many-body localized phase.
We argued that this was well-approximated by quantum facilitated hopping model:
interactions make transitions between Anderson orbitals possible.
We then coupled this quantum facilitated hopping model to a bath.
Crucially, we chose the bath to couple to the system's Anderson orbitals, not to physical sites.
We argued that the the resulting system-bath combination was well-approximated by a classical facilitated hopping model,
and studied the dynamics of that model.

We observed that%
---in contrast to MBL systems coupled to baths by physical sites\cite{Fischer_2016,Levi_2016}---%
the Anderson-coupled system displays a crossover between ``ergodic'' and ``frozen'' regimes.
The disorder strength $h$ and dephasing time $\tau$ act as tuning parameters for this crossover.
The presence of rare thermal bubbles which survive even deep into the frozen phase spoils sharp signatures of the transition in local observables, yielding a smooth crossover between these two phase that is only well defined upon fixing the timescale.
Both thermal and frozen bubbles follow an exponential distribution of sizes in the critical regime, suggesting that the transition is driven not strictly by contiguous bubbles but rather by a resonant backbone of thermal regions and sites.
A clearer transition is observed upon examining the NESS particle current, which shows a transition from diffusion at weak disorder to subdiffusion at strong disorder, consistent with expectations from the MBL transition in the fully quantum model.

Because we couple the bath to Anderson orbitals, transport in our system is entirely determined by the interplay among interactions (in particular the facilitated hopping of Eq.~\eqref{eq:NetworkHam}), resonances, and the bath coupling strength. We speculate that this facilitated hopping mechanism explains the interaction dependence seen in previous studies\cite{Everest_2017}${}^,$\footnote{In this work, the bath is coupled to physical qubits, but ---because the disorder is significantly stronger, and the Anderson orbitals are very tightly localized--- the physics may be qualitatively the same.}.

While the classical model admits a computationally tractable approach for studying the real-time dynamics of a system with a localization-delocalization transition, there may be limitations and potential improvements associated with the numerical method.
Since the facilitated hopping coefficients fall off exponentially with distance, in sufficiently large systems very distant regions may evolve independently of one another.
To this end it may be possible to improve the runtime by some parallelization scheme which separately evolves disjoint segments of the chain simultaneously.
We note also that the derivation of the rates $w_{ij}$ implicitly takes the limit where $\delta t \gg \tau$.
For very fast hopping in ergodic regions this may no longer be an appropriate assumption.
Nonetheless we expect this approach to perform well in the strong-disorder regime.
Despite neglecting accumulated phases and coherence, this classical model captures well the distribution of bubble sizes and the subdiffusive transport expected from the MBL transition.
This may be attributed to the the fact that the numerical method employs the true distribution of link weights $V_{ij}^{(k)}$ obtained from the quantum model while truncating terms which are increasingly irrelevant at strong disorder. Especially, the onset of subdiffusive transport must originate from a change in the distribution of link weights when approaching the MBL transition.

While we motivated the dephasing here as coming from an external bath, we point out that a local dephasing may arise in an isolated system acting as its own bath.
In this case, a uniform dephasing $\tau$ presents a significant simplification over the inhomogeneous spectral line broadening that one should anticipate from ergodic bubbles embedded in a nonergodic phase.
On the one hand, a larger $\tau$ in thermal bubbles would allow these ergodic regions to grow and enlarge the thermal phase.
On the other hand, however, a reduced energy window in the frozen regions would strengthen their role as bottlenecks to transport, potentially allowing for a sharper transition in the transport properties.
The precise manner in which the results presented here are modified when introducing a self-consistent inhomogeneous $\tau$ remains to be studied.

\begin{acknowledgments}
M.B. acknowledges funding via grant DI 1745/2-1 under DFG SPP 1929 GiRyd.
CDW gratefully acknowledge the U.S. Department of Energy (DOE), Office of Science, Office of Advanced Scientific Computing Research (ASCR) Quantum Computing Application Teams program, for support under fieldwork proposal number ERKJ347.
\end{acknowledgments}

\nocite{apsrev41Control}
\bibliographystyle{apsrev4-1}
\bibliography{main}

\appendix
\section{Off-Resonant Hopping}\label{app:off-resonance}

In order to obtain the semi-classical hopping rates $\omega_{ij}$ given in Eq.~\eqref{eq:hopping_rate}, let us consider a simplified instance of our model involving only two sites.
The Hamiltonian is then given by
\begin{equation}
    H_0 = \varepsilon_0 c_0^\dagger c_0 + \varepsilon_1 c_1^\dagger c_1 + \left[V c_0^\dagger c_1 + h.c.\right].
\end{equation}
Now introduce the random fluctuations given in Eq.~\eqref{eq:Hamfluc} and Eq.~\eqref{eq:noise-spec}.
Let us suppose that we have an initial state $\ket{\psi(0)} = \ket{0,1}$ such that a particle initially resides on site $1$.
We are then interested in the probability of finding the particle at site $0$ at time $t$, given by
\begin{equation}
    P_{1 \rightarrow 0}(t) = \llangle \left \vert \bra{1,0} U(t) \ket{0,1} \right\vert^2 \rangll,
\end{equation}
where $U(t)$ is the unitary time-evolution operator.
At leading order in perturbation theory the overlap is just given by
\begin{equation}
    \bra{1,0} U(t) \ket{0,1} \approx V \int_0^t \dif t_1 e^{-i\varphi(t_1)},
\end{equation}
where the phase $\varphi$ is defined as
\begin{equation}
    \varphi(t_1) = \varepsilon_0 (t - t_1) + \varepsilon_1 t_1 + \int_0^{t_1}\dif t' \eta_{1,t'} + \int_{t_1}^t \dif t' \eta_{0,t'}.
\end{equation}
It is now useful to observe that the integral of the noise variables $\eta$ just corresponds to a Wiener process.
To this end, define
\begin{equation}
    \lambda_i(t) = \int_0^t \dif t' \eta_{i,t'}
\end{equation}
such that the phase may be compactly expressed as $\varphi(t_1) = \varepsilon_0 t + \Delta t_1 + \lambda_1(t_1) + \lambda_0(t-t_1)$.
The probability of finding the particle at site 0 then becomes
\begin{equation}
\begin{aligned}
    V^2 \int_0^t \dif t_1 \dif t_2 \int \mathcal{D}[\lambda] P(\lambda) & e^{i\Delta (t_1 - t_2)}e^{i(\lambda_1(t_1) - \lambda_1(t_2))} \\ &\times e^{i(\lambda_0(t-t_1) - \lambda_0(t-t_2))}.
\end{aligned}
\end{equation}
Here we have taken the notation $\int \mathcal{D}[\lambda]P(\lambda)$ for the noise averaging.
Since $\lambda_i$ are Wiener processes, we have a probability distribution
\begin{equation}
    P(\lambda(t_1) - \lambda(t_2)) = \frac{e^{-\frac{\vert \lambda(t_1) - \lambda(t_2)\vert^2}{2\tau^{-1}\vert t_1 - t_2 \vert}}}{\sqrt{2\pi \tau^{-1}\vert t_1 - t_2\vert}}.
\end{equation}
Then integrating over the noise gives a transition probability
\begin{equation}
    P_{1 \rightarrow 0}(t) = V^2 \int_0^t \dif t_1 \dif t_2 e^{i\Delta (t_1 - t_2)}e^{-\vert t_1 - t_2 \vert \tau^{-1}}.
\end{equation}
For $t \gg \tau$ it is straightforward to find the average rate $w_{1\rightarrow 0} = \partial_t P_{1 \rightarrow 0}(t)$ such that
\begin{equation}
    w_{1 \rightarrow 0} = 2 V^2 \frac{\tau}{1 + \Delta^2\tau^2}.
\end{equation}
We take this as the generic form of the classical transitions rates for the Monte Carlo as given in the main text.

\end{document}